\documentclass[final,5p,times]{elsarticle}
\usepackage{amsmath,amssymb,amsfonts,bm,graphicx,dcolumn,setspace}
\usepackage[utf8]{inputenc}

\biboptions{sort&compress}
\newcommand{\M}{\textbf{M}}
\newcommand{\B}{\textbf{B}}
\newcommand{\A}{\textbf{A}}
\newcommand{\W}{\textbf{W}}
\newcommand{\eg}{\textit{e.g.}}

\usepackage{color}

\newcolumntype{d}[1]{D{.}{.}{#1}}
\journal{Thin Solid Films}
\graphicspath{{Figures/}{Figures/Fitness-space/}{Figures/Condition-number-3-FLC/}}

\begin{document}
 
\begin{frontmatter}

\title{Design, optimization and realization of FLC based Stokes polarimeters and Mueller matrix ellipsometer using a genetic algorithm}
\author{Lars Martin S. Aas\corref{cor1}}
\ead{lars.martin.aas@gmail.com}
\author{Daniel G. Sk\aa{}re}
\author{P\aa{}l G. Ellingsen}
\author{Paul Anton Letnes} 
\author{Morten Kildemo}
\address{Department of Physics, The Norwegian University of Science and Technology (NTNU), N-7491 Trondheim, Norway}

\begin{abstract}
The design of broad-band polarimeters with high performance is challenging due to the wavelength dependence of optical components. An efficient Genetic Algorithm (GA) computer code was recently developed in order to design and re-optimize complete broadband Stokes polarimeters and Mueller matrix ellipsometers (MME). Our results are improvements of previous patented designs based on two and three ferroelectric liquid crystals (FLC), and are suited for broad-band hyperspectral imaging, or multichannel spectroscopy applications. We have realized and implemented one design using two FLCs and compare the spectral range and precision with previous designs.
\end{abstract}
\end{frontmatter}

\section{Introduction}
\label{sec:Introduction}

A polarimeter is an instrument that measures the polarization of light to gain additional information compared to what simple intensity measurements reveal. 
By measuring how the polarization of light is altered after being reflected from a surface, the technique is often referred to as ellipsometry.

The need for fast broadband Mueller matrix ellipsometers and Stokes polarimeters result in challenging design problems when using active polarization modulators that are strongly dispersive. Although designs based on \eg{} the Fresnel bi-prism and alike are nearly achromatic, these are not well suited for neither imaging application nor high speed. In the case of polarimeters and Mueller matrix ellipsometers based on liquid crystal modulators, the direct search space may become huge \cite{Letnes:10} and standard optimization methods will evidently result in local minima far away from the optimum. An efficient Genetic Algorithm (GA) computer code was recently developed in order to design and re-optimize complete broadband Stokes polarimeters and Mueller matrix ellipsometers (MME) \cite{Letnes:10}. This code is now used to search systems generating and analyzing optimally selected polarization states, in order to reduce noise propagation to the measured Mueller matrix. Although the GA code was initially 
motivated by the challenging task of searching the components, states and azimuthal orientations for optimally conditioned broadband liquid crystal based polarimeters~\cite{Letnes:10,Aas2013}, the software is written in a versatile manner in order to handle general polarimeters based on any polarization changing components. For small scale production, we also propose that the GA algorithm can be used to  re-optimize the design due to imperfect polarization components, \eg{} due to small deviations in thickness of retarders from manufacturer. Finally, it is noted that the addition of any additional non-trivial polarization altering components in the polarimeter, such as mirrors and prisms also require a re-optimization, which can easily be handled by the GA algorithm, as long as the dispersive properties of such components have been characterized in advance.

We have chosen to use a classical GA \cite{Goldberg:1989uq,Holland:1992kx} to optimize the designs. We present here the optimization of a polarimeter based on Ferroelectric Liquid Crystals (FLC). Such a system was first proposed by Gandorfer~\cite{gandorfer:1402} and Jensen and Peterson~\cite{jensen:42}, and has the advantage of being fast~\cite{Aas} and having no moving parts, which is an advantage for imaging applications. A multichannel spectroscopic Mueller matrix ellipsometer based on this technology is also commercially available (\emph{MM16, Horiba Yvon Jobin}), where this latter system was originally designed for the 430-850 nm wavelength range. The FLC system is based on optical components which are well described~\cite{Ladstein:2008uq,Aas}, but the overall performance of the polarimeter depends on these simple components in a complex manner and traditional optimization routines will be hampered by local minima and the large search space. A genetic optimization algorithm will move out of local 
minima and might find better solutions, resulting in a polarimeter design with less noise amplification on a broader spectral range. 
We here propose several designs, but have as an example here restricted ourselves to only implement small modifications to the commercial MM16 system. Furthermore, we demonstrate how the GA algorithm may be used in small scale production, where we may simply re-optimize the design in the case of an off-specification component. 

\section{Theory}
\label{sec:Theory}
\begin{figure}[htbp]
    \centering 
    \includegraphics[width=0.8\columnwidth]{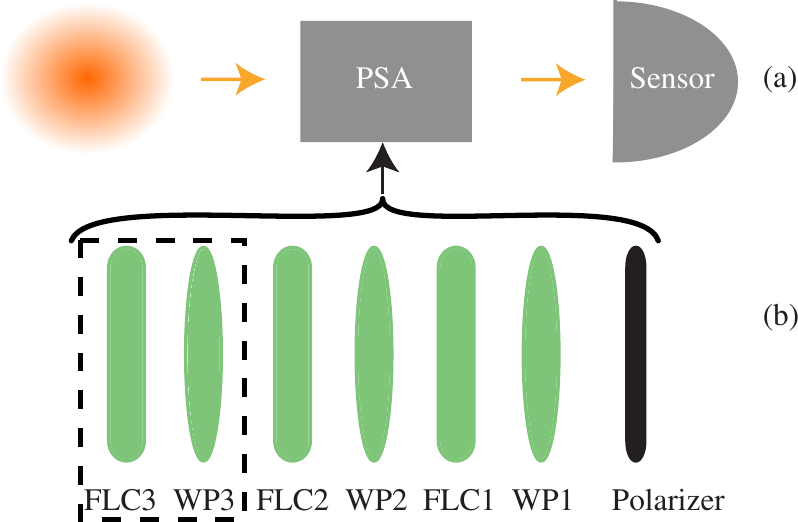}
    \caption{\label{fig:polarimeter-sketch}
	A schematic drawing of a polarimeter, (a) shows a general polarimeter where the polarization state of incident light is analyzed by the Polarization State Analyzer and a light intensity detector. In (b) the components of a Polarization State Analyzer is exemplified through a combination of two or three FLCs and waveplates~(WP) and a linear polarizer.
    }
\end{figure}
The complete polarization state of light, including partially polarized states, can be expressed concisely using the Stokes vector, which completely describes the polarization state with four real elements~\cite{Hauge1980}:
\begin{align*}
    \bm{S} =
    \begin{bmatrix}
        I \\
        Q \\
        U \\
        V \\
    \end{bmatrix}
    =
    \begin{bmatrix}
        \langle E_{0,x}(t)^2\rangle  +  \langle E_{0,y}(t)^2\rangle \\
         \langle E_{0,x}(t)^2\rangle  - \langle E_{0,y}(t)^2\rangle  \\
        2 \langle E_{0,x}(t) E_{0,y}(t) \cos\delta(t)\rangle \\
        2 \langle E_{0,x}(t) E_{0,y}(t) \sin\delta(t)\rangle \\
    \end{bmatrix},
\end{align*}
where the notation $\langle\cdots\rangle$ denotes time average over the in general quadratic time dependent orthogonal electric field components ($E_{0,x}(t)$ and $E_{0,y}(t)$) and phase ($\delta(t)$).

The change of a polarization state can be described by a $4\times4$ real-valued transformation matrix called a Mueller matrix, connecting an incoming Stokes vector $\mathbf{S}_\text{in}$ to an outgoing Stokes vector $\mathbf{S}_\text{out}$,
\begin{equation}
	\mathbf{S}_\text{out}=\mathbf{M} \mathbf{S}_\text{in}.
	\label{eq:Mueller}
\end{equation}
The Mueller matrix can describe the effect of any linear interaction of light with a sample or an optical element. Polarization effects contained in a Mueller matrix could be diattenuation (different amplitude transmittance or reflectance for different polarization modes), retardance (\emph{i.e.} changing $\delta$), and depolarization (which increases the random component of the electric field).  

A Stokes polarimeter consists of a polarization state analyzer (PSA) capable of measuring the Stokes vector of a polarization state, by performing at least four intensity measurements at different projection states. For a given state $(i)$, the polarization altering properties of the PSA can be described by its Mueller matrix $\mathbf{M}^\text{PSA}(i)$, which can be found as the matrix product of the Mueller matrices of all the optical components in the PSA. 
These components are a linear polarizer, and a number of phase retarders (\emph{e.g.} FLCs and waveplates).
An FLC is a phase retarder which can be electronically switched between two states. The difference between the states corresponds to a rotation of the fast axis by $45^\circ$ ($\theta(0)=\theta_0$ and $\theta(1)=\theta_0+45^\circ$). By using a linear polarizer and two FLCs as a PSA, one can generate $2^2 = 4$ different projection states, by using three FLCs one can generate $2^3 = 8$ states, \emph{etc}. 

If an unknown polarization state with Stokes vector $\mathbf{S}$ passes through the PSA, for a given projection state $i$, as given in Equation~\eqref{eq:Mueller} the detector will measure an intensity $I$ depending only on the first row of $\mathbf{M}^\text{PSA}(i)$,
\begin{equation*}
	 I=\sum_{j=1}^{4} \mathbf{M}^\text{PSA}_{1,j}(i) \mathbf{S}_j,
	\label{eq:}
\end{equation*}
 and can be considered to be a projection of $\mathbf{S}$ along a Stokes vector equal to $\mathbf{M}^\text{PSA}_{1,1..4}~^T$, where $T$ denotes the transpose. These Stokes vectors are organized as rows in the system matrix $\mathbf{A}$, which when operated on a Stokes vector gives
 \begin{equation*}
	\mathbf{b}=\mathbf{A}\mathbf{S}, 
	\label{eq:}
\end{equation*}
where $\mathbf b$ is a vector containing the intensity measurements at the different projection states. The unknown Stokes  vector can then be found by inverting $\mathbf{A}$, $\mathbf{S}=\mathbf{A}^{-1}\mathbf{b}$. The noise in the intensity measurements $\mathbf{b}$ will be amplified by the condition number~($\kappa$) of $\mathbf{A}$ in the inversion to find $\mathbf{S}$~\cite{NumRes}. Therefore $\kappa$ of a polarimeter should be as small as possible~\cite{Tyo2000}, which correspond to doing as independent measurements as possible (\emph{i.e.} to use projection states that are as orthogonal as possible). The condition number of $\mathbf{A}$ is given as $\kappa=\|\mathbf{A}\|\|\mathbf{A}^{-1}\|$, which for the 2-norm is equal to the ratio of the largest to the smallest singular value of the matrix~\cite{NumRes}. The best condition number that can be achieved for a polarimeter is $\kappa=\sqrt3$~\cite{Tyo2000}. If four optimal states can be achieved, no advantage is found by doing a larger number of 
measurements with different states, compared to repeated measurements with the four optimal states~\cite{Sabatke2000}. If, however, these optimal states can not be produced ($\kappa>\sqrt3$), the condition number, and hence the error, can be reduced by measuring more than four states. For an FLC based polarimeter this can be done by using three FLCs followed by a polarizer as the PSA, with up to three waveplates (WP) coupled to the FLCs to reduce the condition number (see Figure~ \ref{fig:polarimeter-sketch}), or components with more than two states, such as liquid crystal variable retarders (LCVR). In this case $\mathbf{A}$ will not be a square matrix, and the \emph{Moore-Penrose pseudoinverse} is then used to invert $\mathbf{A}$~\cite{Aas2013}. 

To measure a Mueller matrix of a sample, it is necessary to illuminate the sample with at least four different polarization states. The Stokes vectors of these states can be organized as columns in a polarization state generator (PSG) system matrix $\mathbf{W}$. The product $\M\W$ gives the resulting four Stokes vectors after interaction with the sample, they are then measured by the PSA, yielding the intensity matrix $\B=\A\M\W$. The Mueller matrix can then be found by multiplying the expression by $\A^{-1}$ and $\W^{-1}$ from each side, $\M=\A^{-1}\B\W^{-1}$. For overdetermined PSA and PSG $\A$ and $\W$ are not square, the More-Penrose pseudo-inverse can then be used to find the best inverse. The PSG may be constructed from the same optical components and has the same optimum configuration as the PSA.

\section{Fitness evaluation}
\label{sec:Fitness evaluation}
We have already established that $\kappa$ should be as small as possible in order to reduce noise in the polarimetric measurements. It is a fairly trivial exercise to optimize $\kappa$ for a single wavelength. However, there are two sources of wavelength dependence of the optical properties of the components.

One of these is the explicit wavelength dependence of the retardance $\Delta_R$, which can be calculated as~\cite{hecht}
\begin{align}
\label{eq:retardance}
    \Delta_R = \frac{2\pi l ( \Delta n)}{\lambda_0},
\end{align}
where $l$ is the physical thickness of the component (\emph{e.g.} waveplate or FLC), $\lambda_0$ is the vacuum wavelength of the light, and $\Delta n$ is the birefringence of the material. Birefringence is the difference in refractive index between the fast axis (index of refraction $n_\text{f}$) and the slow axis ($n_\text{s}$), \emph{i.e.} $\Delta n = |n_\text{f} - n_\text{s}|$~\cite{hecht}. There is an explicit wavelength dependence in Eq.~\eqref{eq:retardance}, which complicates the design of the PSA. A weaker but still important effect is the wavelength dependence of the birefringence, \emph{i.e.} $\Delta n = \Delta n (\lambda)$. Both of these effects are taken into account by using experimental data for the retardance~\cite{Ladstein2007, Aas2013}.

To evaluate the performance of a polarimeter design, we compare the inverse condition number ($\kappa^{-1}(\lambda)$) to the theoretically optimal value ($1 / \sqrt{3}$). The argument for using $\kappa^{-1}$ rather than $\kappa$ is that $\sqrt{3} < \kappa < \infty$ while $0 < \kappa^{-1} < 1/\sqrt{3}$; the latter range is more numerically convenient. In detail, we define an ``error function'' ($e$) as
\begin{align}
    \label{eq:error-function}
    e = \frac{1}{N_\lambda} \sum_{n=1}^{N_\lambda}
        \left(
            \kappa^{-1}(\lambda_n) - 1/\sqrt{3}
        \right)^4.
\end{align}
In the above equation, we typically use $\lambda_n = \lambda_\text{min} + (n - 1) \Delta\lambda$, with $n = 1, 2, \ldots, N_\lambda$ and $\Delta\lambda = 5$ nm. It is, of course, possible to choose other discretization schemes for $\lambda$:  for some applications, one can \emph{e.g.} be interested in optimal performance near a few spectral lines (wavelengths). We take $\left(\kappa^{-1} - 1 / \sqrt{3} \right)$ to the power of four to punish unwanted peaks in $\kappa$ more severely. As GAs conventionally seek to maximize the fitness function, we define our fitness function $f$ as
\begin{align}
    \label{eq:fitness-function}
    f = \frac{1}{e}.
\end{align}
As $e$ will never be zero in practice, there is no need to add a constant term in the denominator. The fitness function does not carry any physical significance on its own; it is simply an overall measure of how well the polarimeter can measure along orthogonal polarization states for the chosen wavelengths.

\section{Genetic algorithm}
\label{sec:Genetic algorithm}
The GA was based on the open Python library Pyevolve~\cite{ref:Pyevolve}, and was written to handle any kind of optical components. We have however concentrated our efforts on systems based on liquid crystals (and in particular FLCs) as polarization modulators, with fixed waveplates ``sandwiched'' between them, this coupling of waveplates and FLCs enables the achromatic design. The designs we set out to find was one based on three FLC retarders and three fixed waveplates, and one based on two FLC retarders and two fixed waveplates. Each FLC has two variables, namely the normalized thickness $L$ (defined in Eq.~\eqref{eq:retardance-formula}, influencing the retardance $\Delta_R$) and its orientation angle $\theta$. The same is true for the fixed waveplates. This yields 12- and 8-dimensional search spaces: six and four components with two variables each.

\label{sub:Representation}
In the GA, polarimeter designs are represented using a traditional binary genome. Each component is assigned a number of bits for $\theta$ and a number of bits for $L$. $\theta$ is the simplest case, as its possible values are limited: the best achievable alignment accuracy is estimated to $\Delta\theta \approx 0.5^\circ$, and $0 < \theta^\circ < 180^\circ$. This means that $8$ bits per component for the variable $\theta$ is sufficient. For $L$, one should choose a minimum and maximum value according to which components can be realistically purchased. Here, too, is the experimental resolution somewhat coarse, so that one does not need a large number of bits for its representation ($8$-$10$ bits is sufficient). After determining $L$ and $\theta$ for each of the six or four components, we proceed by determining the full transfer matrix of the PSA, $\mathbf{M}^\text{PSA}(\lambda,i)$ for each discrete wavelength $\lambda_n$ and each projection state $i$. As described in Section~\ref{sec:Theory}, one can 
determine the condition number $\kappa(\lambda)$ for $\mathbf{A}$ from the transfer matrices $\mathbf{M}^\text{PSA} (\lambda, i)$. The first generation was initialized by generating genomes with the bits chosen randomly with a uniform distribution.

Initially, we let component ordering be a variable in our genome. In that case, the first few bits of the genome would determine the ordering of the components. This was done by interpreting these bits as the index in a list of components. However, the best results from initial simulation runs almost always had the same component ordering as older ``non-genetic'' designs~\cite{Aas, Ladstein2007, GarciaCaurel2004}. Hence, we removed this feature to speed up convergence.

\label{sub:Genetic operators}
The genetic operators that were used are the well known ones for binary genomes~\cite{Goldberg:1989uq,Holland:1992kx}. For mutation, the simple bit-flip operator was used; \emph{i.e.} flipping $0\rightarrow 1$ or vice versa. The mutation rate per individual was typically set to $0.2$ per generation. Crossover was performed by multi-point crossover. Experience indicates that two crossover points combined with a crossover rate of $0.7$ gives the best convergence performance. The selection protocol we used was tournament selection with $K = 4$ individuals in the tournament pool and $\varepsilon = 0.3$ probability of an ``underdog'' selection. The elitism rate was set to $1$ individual per generation. It should be noted that depending on the number of components and, hence, the genome length, the exact rates may have to be adjusted somewhat for optimal performance.

In the final simulations a population of $500$ individuals evolved over $600$ generation. Several equivalent simulation runs were performed with different initializations of the random number generator. As the theoretically optimal performance for realistic materials is not known, no other convergence criteria than the maximum number of generations was used. Decent results can, however, be achieved more quickly with smaller population sizes and a lower number of generations.

\section{Results}
\label{sec:Results}
While the GA can handle components with arbitrary dispersion relations, we limit our discussion to components whose wavelength-dependent retardance can be fitted to the following modified Sellmaier equation:
\begin{align}
    \label{eq:retardance-formula}
    \Delta(\lambda) \approx 2\pi L
        \left[
            \frac{A_{UV}}
                {(\lambda^2 - \lambda_{UV}^2)^{1/2}}
            - \frac{A_{IR}}
                {(\lambda_{IR}^2 - \lambda^2)^{1/2}}
        \right]
\end{align}
Here, we call $L$ the normalized thickness of the component, as $L$ is proportional to the component's physical thickness. The parameters $A_{UV}$, $A_{IR}$, $\lambda_{UV}$, and $\lambda_{IR}$ can be found by fitting experimental data to this model. Initially, for the results presented in this paper, numerical values from characterization experiments~\cite{Ladstein2007} performed on quartz waveplates and FLCs were used. After the initial optimization FLCs with the optimal thickness were ordered from \textit{Citizen}. After receiving the components, new thickness characterizations were made, and the system was re-optimized on waveplate thickness and orientation, and FLC orientation. In the end, the final orientation of all components were found after characterization of the thicknesses. As a result, the PSA and PSG do not have the same design, due to differences caused by the optical component manufacturing precision.   

We seek an improved design for the commercially available MME using two FLCs by having a lower condition number over a larger spectral range. As 1000 nm is typical the upper wavelength limit for a silicon detector based spectrograph the range in the optimization was set from 430 nm to 1000 nm, an improvement from 850 nm compared to the commercial system. We also aimed at lower noise amplification in the whole spectral range. 

\begin{figure}[htb]
\begin{center}
  \includegraphics[width=0.9\columnwidth]{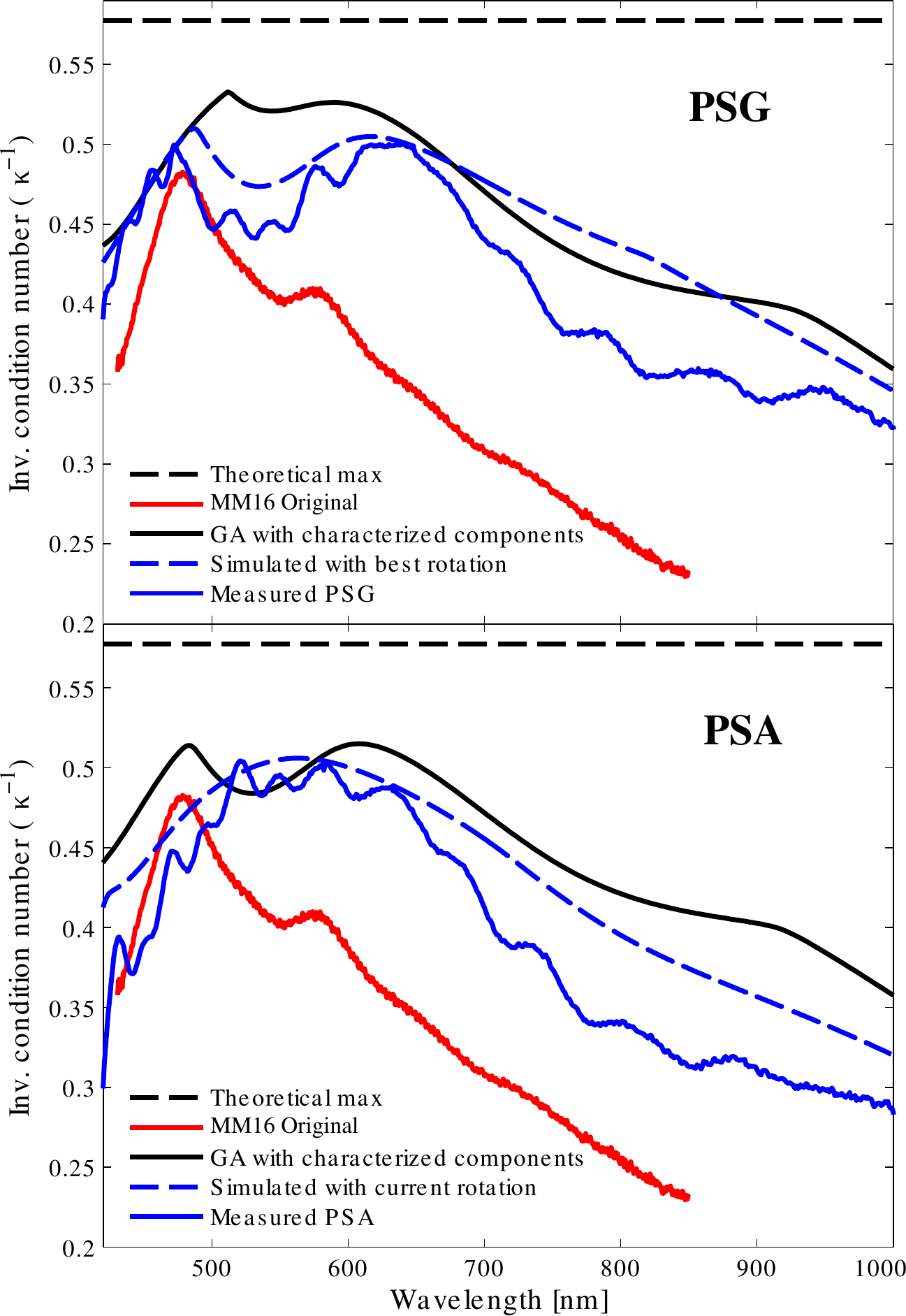}
  \caption{Inverse condition number for polarimeter designs based on two FLCs and two waveplates. Results are shown for an older design in the visible~\cite{mm16}. We also show GA generated designs that cover a wider spectral range with a better condition number for optimal theoretical design (solid black), best design after characterization (dashed blue) and best achieved design after mounting and calibration (solid blue).}
  \label{fig:condPSGPSA}
\end{center}
\end{figure}
The resulting condition number from the optimized polarimeter using two FLCs are shown in Figure~\ref{fig:condPSGPSA}. It is noted that the system design was somewhat limited by the fact that thin FLCs could not be currently manufactured. Only a quasi-optimal system with given FLC manufacturer limitations could be designed. The results from the PSG and the PSA are plotted separately. In both axes the dashed black line shows the theoretical best inverse condition number, the solid red shows the measured inverse condition number on the commercial MM16 instrument. The solid black curve shows the inverse condition number for the optimal configuration using the measured dispersion of the individual components, the dashed blue curve shows the simulated inverse condition number using the actual experimental orientations, and finally, the solid blue curve shows the actual measured, and calibrated inverse condition number for the final designs. All designs are better conditioned than the previously designed 
instrument, and allows for measurements of the Mueller matrix across a broader spectral range. The system can operate down towards the cut-off of the silicon spectrograph detector. Some interesting issues appeared in the implementation of the PSA/PSG, which is related to the mounting accuracy of the optical components. It is noted that certain designs may be more sensitive to small azimuthal mounting errors. One could thus envisage to include in the fitness function certain mounting inaccuracies in order to also search for the most robust design for larger scale production. 

\begin{figure}[tbhp]
    \centering
    \includegraphics{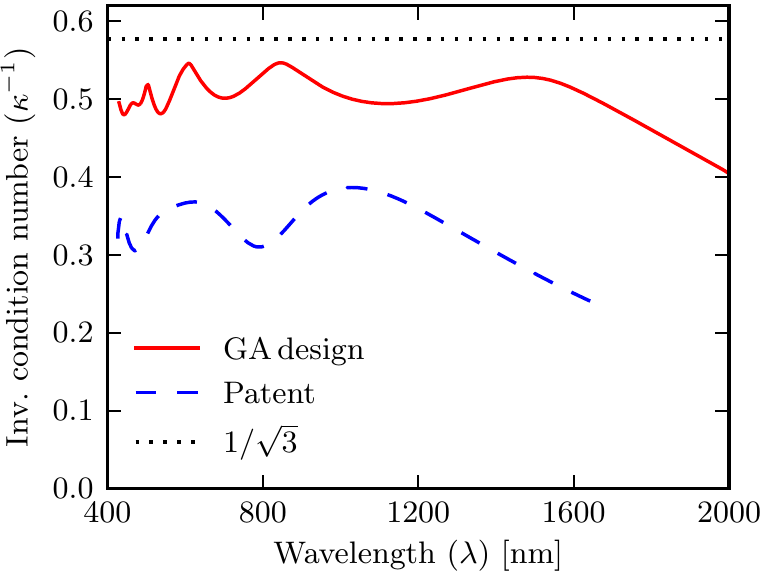}
    \caption{\label{fig:condition-numbers}
        Inverse condition number ($\kappa^{-1}(\lambda)$) for a GA-generated and a previously patented design~\cite{Patent3FLC}. The GA-generated design is based on three FLCs and three waveplates, while the previous patented design is based on three FLCs and one waveplate. The GA design is significantly better than the previous design for all wavelengths. If $\kappa^{-1} \lesssim 0.2$, noise becomes very problematic, and the instrument is considered unreliable. $\kappa^{-1}$ is theoretically limited to the range $0 < \kappa^{-1} < 1/\sqrt{3}$.
    }
\end{figure}

We also briefly recall that we have recently reported systems designed using three FLCs in the PSG and PSA for an extended wavelength range from 430 nm to 2000 nm \citep{Letnes:10}. Here, the power of the GA design algorithm becomes even more evident, which is clearly seen by the polarimeter design shown in Figure~\ref{fig:condition-numbers}. The red solid line shows $\kappa^{-1}(\lambda)$ which is our measure of performance, where a higher value of $\kappa^{-1}$ is better. The design parameters of the polarimeters, \emph{i.e.} the $\theta$ and $L$ values, for both the three and two FLC design are shown in Table~\ref{tab:optimal-flc-polarimeter}. For comparison with previous designs, we show a recently patented design~\cite{Patent3FLC} in comparison with the GA generated one. The GA generated design is based on three FLCs and three waveplates, while the previous patented design is based on three FLCs and one waveplate. The new design is useful over a broader spectral range (here defined as the parts of the spectrum where $\kappa^{-1} \gtrsim 0.2$) and has lower 
noise amplification due to a lower condition number (higher inverse condition number). It should be noted that the FLC technology is limited downwards in wavelength to $430$ nm, because of material degradation from ultra violet light.

\begin{table*}[hbt]
    \centering
    \caption{Orientation angle, $\theta$, and normalized thickness, $L$, for all components of the best polarimeter based on three and two FLCs, as shown in Figures~\ref{fig:condition-numbers} and~\ref{fig:condPSGPSA}. For comparison, we also include the wavelength where the retardance is $\lambda/4$ (for some components, for $\lambda/2$) for our design as well as the patented design with three FLCs. The notation WP1, FLC1 \emph{etc.} is explained in Figure~\ref{fig:polarimeter-sketch}. Note that the previous patented design uses only one fixed waveplate, while our design uses three.}
    \vspace{0.15cm}
    \begin{tabular}{|l|d{-1}d{-1}r|d{-1}c|d{-1}d{-1}r|}
    \hline
                & \multicolumn{3}{c|}{Three FLC design}& \multicolumn{2}{c|}{Three FLC Patent} &\multicolumn{3}{c|}{\text{2 FLC Visible design}}\\
    Component   & \multicolumn{1}{c}{$\theta [^\circ]$} & \multicolumn{1}{c}{$L$} & \multicolumn{1}{c|}{$\lambda/ 4$ @} & \multicolumn{1}{c}{$\theta [^\circ]$} & \multicolumn{1}{c|}{$\lambda/ 4$ @} & \multicolumn{1}{c}{$\theta [^\circ]$} & \multicolumn{1}{c}{$L$} & \multicolumn{1}{c|}{$\lambda/ 4$ @}\\
    \hline
    FLC1    & 56.5  & 2.44 &  1991 nm & 46.0 & 1150  nm & 100.6 & 1.06 & 894\text{ nm}\\
    WP1     & 172.9 & 1.10 &  493 nm & &  &  10.2 & 3.37 & 1404\text{ nm}\\
    FLC2    & 143.3 & 1.20 &  1009 nm & -5.0 &  1050  nm & 89.9 & 1.05 & 901\text{ nm} \\
    WP2     & 127.1 & 1.66 &  722 nm & 92.0 & $\lambda/2$ (Achromatic) & 18.5 & 3.75 & 1552\text{ nm}\\
    FLC3    & 169.4 & 1.42 &  1181 nm & 72.0 &  ~600  nm & & & \\
    WP3     & 110.1 & 4.40 &  1798 nm & & & & &\\
    \hline
    \end{tabular}
    \label{tab:optimal-flc-polarimeter}
\end{table*}

\begin{figure}[bht]
    \centering
    \includegraphics{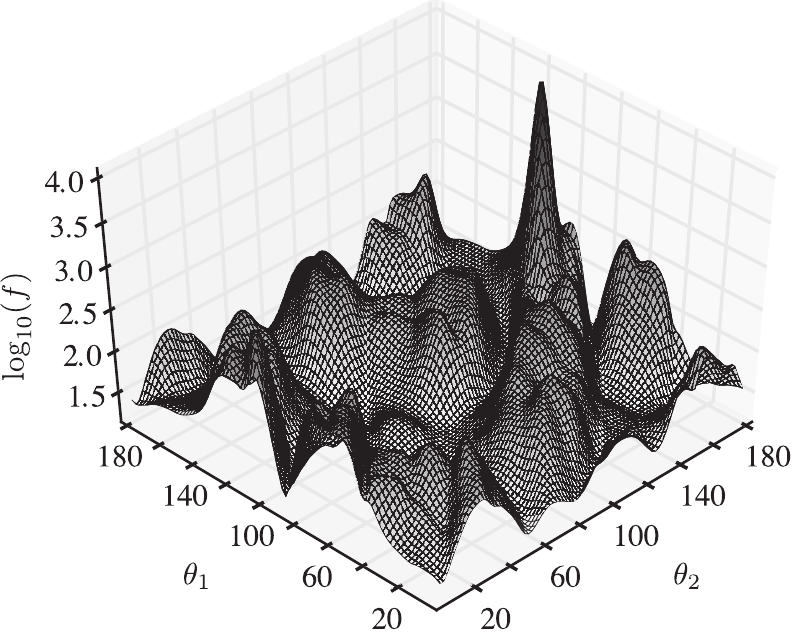}
    \caption{A two-dimensional ``cut'' of the fitness landscape around the optimal value for the GA-generated design with three FLCs shown in Figure~\ref{fig:condition-numbers}. $\theta_1$ is the orientation angle of FLC3 and $\theta_2$ is the orientation angle of WP3, as shown in Figure~\ref{fig:polarimeter-sketch}. The other $\theta$ and $L$ parameters were set to the optimal values.}
    \label{fig:fitness-landscape}
\end{figure}
One can get an impression of how complex the fitness landscape is from Figure~\ref{fig:fitness-landscape}. Here, a plot of $f(\theta_1, \theta_2)$ is shown, where $\theta_1$ is the orientation angle of FLC3 and $\theta_2$ is the orientation angle of WP3, the two first components in Figure~\ref{fig:polarimeter-sketch}. All other parameters, \emph{i.e.} $\theta$ and $L$ values for the other components, were set to the optimal value as given in Table~\ref{tab:optimal-flc-polarimeter}. Note that $f(\theta_1, \theta_2)$ is periodic in both variables with a period of $180^\circ$. Due to the enormous number of local minima, even in only $2$ of the $12$ search dimensions, a clever optimization algorithm is required.

\begin{figure*}[htb] 
\begin{center}
\includegraphics[width=14 cm]{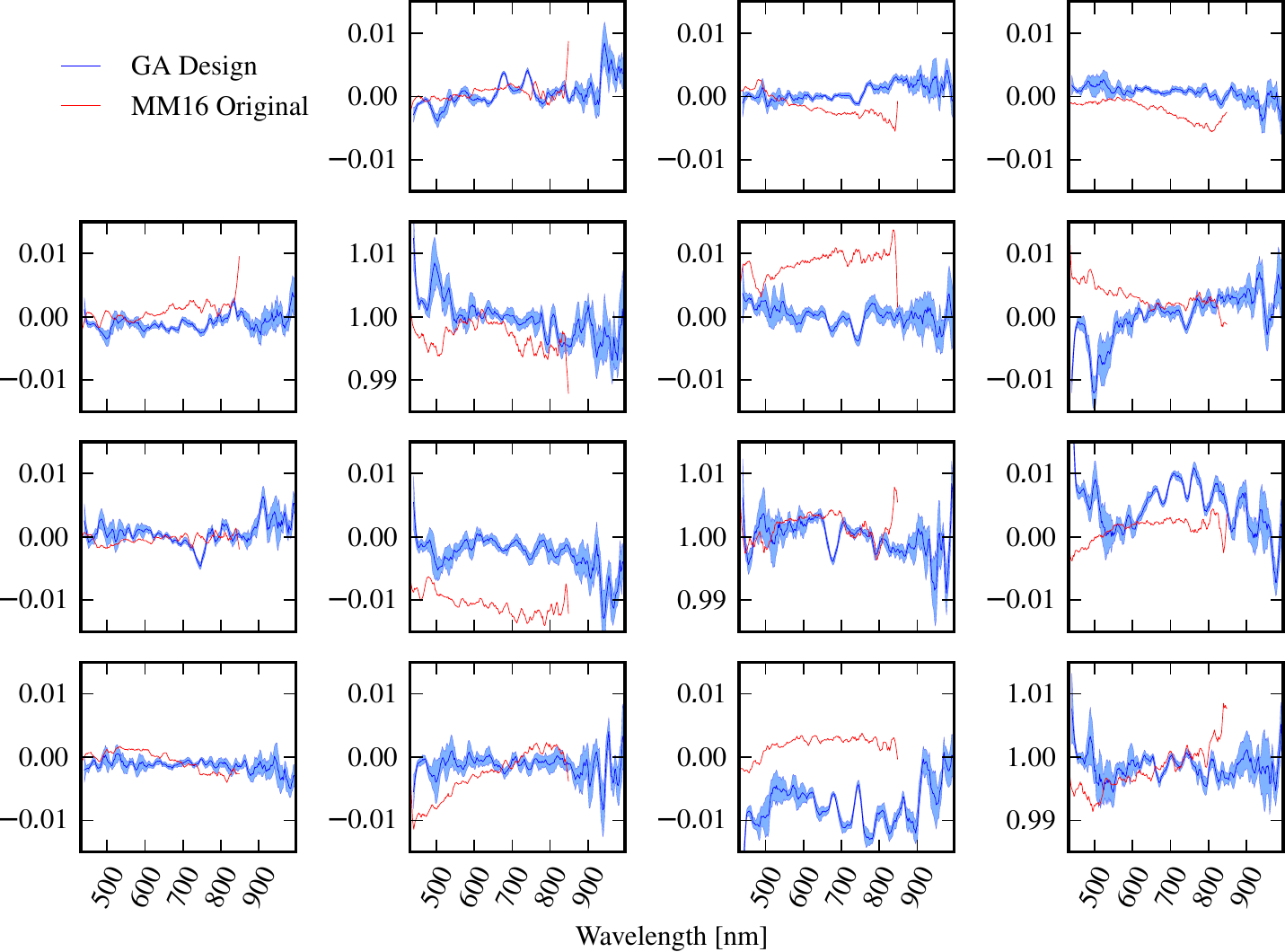}
  \caption{Mueller matrix measurement of air (identity matrix) including the measurement using the original vis-design (red) and the measurement using the new design (blue), the light blue area shows the standard deviation of 10 measurements. The new design allows the wavelength range to be extended to 1000 nm, while the precision of the new and old design appears to be similar.}
  \label{fig:MMair}
  \end{center}
\end{figure*}
The Mueller matrix ellipsometer based on two FLCs in PSG and PSA (Figure~\ref{fig:condPSGPSA}) where inserted into the MM16 instrument from Horiba and calibrated the normal way using the eigenvalue calibration method~\cite{Compain:99} implemented in the software DeltaPsi 2 from Horiba. To verify the precision of the instrument, ten measurements of air were made, these are plotted in Figure~\ref{fig:MMair} together with a measurement using the old design (red curve). The mean of the ten measurements are plotted with a dark blue curve and the standard deviation is plotted as the light blue area around the curve. There is no evident difference in accuracy or precision between the measurements using the two different designs. The maximum error is approximately one per cent. The obvious improvement is the operation across an enlarged spectrum.  

An important application of FLC based polarimeters are in addition to the spectroscopic ellipsometry the Mueller matrix imaging~\cite{Hatit2008,Ellingsen2011,Aas2010}, where the increased bandwidth allows variations important in particular for biological imaging.

\section{Conclusion}
\label{sec:Conclusion}
Genetic Algorithms (GA) are able to generate optimized designs of Stokes/Mueller polarimeters covering a broader spectral range with reduced noise amplification (lower system matrix condition numbers). Compared to previous optimization techniques used for this purpose often based on direct or gradient searches in small parts of the search space, the GA outperforms these methods when having multidimensional search spaces with many local minima. 
An instrument based on ferroelectric liquid crystal retarders optimized using the GA was assembled and characterized showing system properties as expected from the simulations, with extended spectral range.
\section*{Acknowledgements}
\label{sec:Acknowledgements}
LMSA acknowledges financial support from The Norwegian Research Center for Solar Cell Technology (project num. 193829). The authors are grateful to Denis Catelan, Horiba for assisting and advising on the mechanical parts, and the extension of the spectrograph range of the MM16.
 
%  \bibliographystyle{model1a-num-names}
%  \bibliography{references}

\end{document}